\documentclass[onecolumn]{article}
\usepackage{graphicx} 
\graphicspath{{figures/}}
\usepackage{subcaption}
\usepackage{multirow}
\usepackage{multicol}
\usepackage[margin=2.5cm]{geometry}
\usepackage{amsfonts,amsmath,amssymb}
\usepackage{booktabs}
\usepackage{authblk}
\usepackage{todonotes}
\usepackage{enumerate}
\usepackage{array} 
\usepackage{url} 
\usepackage{hyperref}
\newcommand{\ket}[1]{\ensuremath{\left|#1\right\rangle}} 

\newcommand{\revfix}[1]{{1}}
\let\vec\mathbf

\usepackage{parskip}
\usepackage{titling}

\title{Dimension reduction with structure-aware quantum circuits 
for hybrid machine learning}
\author{Ammar Daskin}
\affil{
Department of Computer Engineering\\
Istanbul Medeniyet University\\ 
Istanbul, Turkiye, 34000\\
Email Address: adaskin25@gmail.com\\
Orcid ID: \href{https://orcid.org/0000-0002-1497-5031}{0000-0002-1497-5031}
}
\date{
}
\begin{document}

\maketitle

\begin{abstract}
Schmidt decomposition of a vector can be understood as writing the singular value decomposition (SVD) in vector form. A vector can be written as a linear combination of tensor product of two dimensional vectors by recursively applying  Schmidt decompositions via SVD to all subsystems.
Given a vector expressed as a linear combination of tensor products, using only the $k$ principal terms yields a $k$-rank approximation of the vector. Therefore, writing a vector in this reduced form  allows to retain most important parts of the vector while removing small noises from it, analogous to SVD-based denoising.

In this paper, we show that quantum circuits designed based on a value $k$ (determined from the tensor network decomposition of the mean vector of the training sample) can approximate the reduced-form representations of entire datasets. We then employ this circuit ansatz with a classical neural network head to construct a hybrid machine learning model. Since the output of the quantum circuit for an $2^n$ dimensional vector is an $n$ dimensional probability vector, this provides an exponential compression of the input and potentially can reduce the number of learnable parameters for training large-scale models.
We use datasets provided in the Python scikit-learn module for the experiments. The results confirm the quantum circuit is able to compress data successfully to provide effective $k$-rank approximations to the classical processing component.
    \textbf{\textit{Keywords: Quantum dimension reduction; quantum machine learning; Schmidt decomposition; quantum circuits} }
\end{abstract}

\section{Introduction}
\subsection{Motivation}
\textbf{Reducing number of parameters in machine learning models.} Pruning techniques \cite{vadera2022methods} have been shown to dramatically decrease the number of parameters in trained neural networks \cite{denil2013predicting,marion2023less,frankle2018lottery}. Although pruning  decreases the trained model sizes and enables the trained models to be deployed on small machines \cite{zhu2017prune}, it does not necessarily address the resource requirements for training high-dimensional data which may be containing many redundancies.
The dimension of the data can be reduced by using principal component analysis through applying singular value decomposition to the data. However, although this removes the redundancies from the data, it does not dramatically reduces the data dimension and typically provides only polynomial reduction. Recent studies (e.g. \cite{huang2021power,cerezo2021variational,havlivcek2019supervised,daskin2025learnable}) suggest quantum computers may enable exponential reduction of data dimensionality in machine learning and therefore can provide a way to reduce required resources \cite{cerezo2022challenges}: This mainly stems from quantum frameworks where:  while the input given the quantum framework is an entire $2^n$-dimensional vector (quantum state for $n$ qubits), the output is an $n$-dimensional probability vector (input to classical neural network). Specifically, such frameworks provide mappings from $2^n$-dimensional spaces to $n$-dimensional probability spaces using quantum circuits with only $\text{poly}(n)$ parameters.
Therefore, the resulting hybrid models require only polynomially many parameters in the number of qubits. This simplifies training and reduces the need for extensive pruning since the resulting model is already a considerably small model.

\textbf{Tensor network decomposition of a vector.} 
The singular value decomposition of a matrix expressed in vector form gives a linear combination of Kronecker products between the left and right singular vectors. In quantum computing, this representation corresponds to the Schmidt decomposition of a quantum state, where the two subsystems are spanned by the left and right singular vectors. The singular values serve as coefficients that quantify the entanglement between these subsystems. 
Using successive Schmidt decomposition, an arbitrary vector can be expressed as a linear combination of Kronecker tensor products: $\ket{\psi}=\sum_i s_i \bigotimes_j^n \vec{u}_{ij}$, where $\vec{u}_{ij}$ is is a normalized two-dimensional vector.
Analogous to obtaining $k$-rank matrix approximations via SVD, a data vector $\ket{\psi}$ can be approximated by its reduced form $\ket{\psi_k} = \sum_{i=1}^k s_i \bigotimes_{j=1}^n \vec{u}_{ij}$, which preserves the most significant features of the originanl vector.
Quantum computers can efficiently process vectors represented as single tensor products (unentangled states) without extensive compilation. Furthermore, linear combinations of such products can be implemented via block encoding techniques \cite{childs2012hamiltonian,daskin2012universal}: This approach primarily uses an ancilla register to control the application of each term to the data qubits, along with an ancilla state-preparation circuit to implement the coefficients.

\textbf{Barren plateaus and training difficulty in quantum machine learning.} Quantum circuits with $n$ qubits can span a Hilbert space of dimension $2^n$. While this provides powerful expressivity for quantum machine learning models based on parameterized quantum circuits, it impedes training when the spanned space becomes excessively large. Specifically, the gradient direction becomes difficult to determine accurately—a phenomenon analogous to vanishing gradients in classical machine learning, known as the barren plateau problem \cite{mcclean2018barren} (see the recent review in Ref.~\cite{larocca2025barren}).
The tensor network decomposition of a network obtained through successive Schmidt decomposition provides an insight into dataset structure. Therefore, it can be used to determine the minimum number of terms required for a faithful representation of the data, which can prevent over-parameterizing quantum circuits. 

\subsection{Contribution}
For a dataset represented by vectors $\{\vec{x}_1, \vec{x}_2, \dots\}$, in this paper we show that a quantum circuit $U$ implementing a generic $k$-term decomposition (i.e., a linear combination of tensor products of two-dimensional vectors) can generate $k$-rank approximations of the data vectors. Specifically, 
\begin{equation}
\{U\vec{x}_1, U\vec{x}_2, U\vec{x}_3, \dots\} \approx \{\vec{x}_1^{(k)}, \vec{x}_2^{(k)}, \vec{x}_3^{(k)}, \dots\}.
\end{equation}
This process is equivalent to reducing the dimension of a data vector through singular value transformation. However, in quantum case, since the circuit outputs $n$-dimensional probability vectors (from $n$ qubits), the crucial point is that this enables exponential compression of the original $2^n$-dimensional data while preserving its most significant features. 

The key contributions of the paper are:
\begin{itemize}
    \item We demonstrate, using example datasets, that after computing the tensor network decomposition of a the mean vector of sample, a quantum circuit implementing a generic $k$-term decomposition can successfully learn to generate the reduced vectors for the whole data set $\{\vec{x}_1^{(k)}, \vec{x}_2^{(k)}, \vec{x}_3^{(k)}, \dots\}$.
    \item We integrate this quantum circuit with a classical neural network head to construct a hybrid machine learning model where the quantum circuit component  compresses $2^n$-dimensional inputs exponentially into $n$-dimensional probability vectors for the classical component.
    \item We show that the hybrid model achieves results comparable to (and in some cases superior to) classical neural networks applied directly to the $2^n$-dimensional reduced vectors.
    \item This framework potentially provides a way to reduce the number of parameters needed to train large-scale datasets.
\end{itemize}

\subsection{The paper outline}
The rest of the paper is organized as follows: Section \ref{sec:background} provides a brief background on singular value decomposition, Schmidt decomposition, and tensor network decomposition,  and shows how they are related. This section also reviews related works in quantum machine learning and dimension reduction. Section \ref{sec:methods} details our proposed method, including the quantum circuit design for implementing $k$-term approximations and its integration into hybrid machine learning frameworks. Section \ref{sec:experiments} presents experimental results on datasets demonstrating: 
\begin{enumerate}[i)]
    \item the approximability of real-world datasets,
    \item quantum circuit performance for $k$-rank approximation, 
    \item and comparisons between hybrid models and classical baselines.
\end{enumerate}
 Finally, Section \ref{sec:discussion} discusses some implications and limitations, and gives some future research directions. Additional sections cover data availability (with a link to the code repository), funding acknowledgments, and conflicts of interest.
\section{Background and related works}
\label{sec:background}
\subsection{Review of SVD, Schmidt, and tensor decomposition}
The singular value decomposition for an $N=2^n$-dimensional matrix is defined as:
\begin{equation}
    A = U\Sigma V^T = \sum_{i=1}^N \sigma_i \vec{u}_i\vec{v}_i^T ,
\end{equation}
where $U$ and $V$ are left and right unitary matrices with columns $\vec{u}_i$ and $\vec{v}_i$. $\Sigma$ is a diagonal matrix of singular values (diagonal entries) $\sigma_1 \geq \dots \geq \sigma_N \geq 0$.
One can obtain a rank-$k$ approximation $A^{(k)}$ of the matrix $A$ by  considering only the largest $k$ singular values $\sigma_i$ in the summation:
\begin{equation}
   {A}^{(k)}= \sum_{i=1}^k \sigma_i \vec{u}_i\vec{v}_i^T.
\end{equation}
The approximation error in the spectral or the Frobenius norm, $\|A-{A}^{(k)}\|$, is then determined by $\sigma_{k+1}$.

We can write this decomposition in vectorized form using the Kronecker tensor product:
\begin{equation}
\text{vec}(A) =  \sum_{i=1}^N \sigma_i \vec{v}_i \otimes \vec{u}_i.
\end{equation}
When a quantum state $\ket{\psi}$ is expressed in the above form, it is called the Schmidt decomposition of the state with respect to a bipartition where the subspaces are represented by the vectors $\vec{v}_i$ and $\vec{u}_i$. In this form, having more than one non-zero Schmidt coefficient $\sigma_i$ indicates entanglement between the subsystems.

Here, both subsystem sizes are equal; therefore, we have the same number of qubits in each subsystem. For an $n$-qubit state $\ket{\psi}$ and $m+l=n$, subsystems with $m$ and $l$ qubits can be analyzed similarly by reshaping $\ket{\psi}$ into a matrix of dimension $2^m \times 2^l$ and then finding its SVD.
For $m=1$, the decomposition gives at most $\min(2, 2^l)$ non-zero Schmidt coefficients:
\begin{equation}
\ket{\psi} =  \sum_{i=1}^{2} \sigma_i \ket{u_i} \otimes \ket{v_i},
\end{equation}
where  $\ket{u_i}$ and $\ket{v_i}$ are orthonormal vectors. While the $\ket{u_i}$ (being single-qubit states) cannot be further decomposed into smaller subsystems, we can recursively decompose the $\ket{v_i}$ (which represent $l$-qubit states). Recursively applying Schmidt decomposition \cite{daskin2023dimension} to subsystems with dimension larger than 2 or running numerical algorithms such as tensor train decomposition \cite{oseledets2011tensor} on the original vector, we can obtain a tensor network decomposition of the state:
\begin{equation}
\ket{\psi} =  \sum_{i_1,\dots,i_n} c_{i_1\dots i_n} \ket{t_{i_1}} \otimes \dots \otimes \ket{t_{i_n}} =  \sum_{i_1,\dots,i_n} c_{i_1\dots i_n} \bigotimes_{j=1}^n \ket{t_{i_j}},
\end{equation}
where the $\ket{t_{i_j}}$ are single-qubit states (basis vectors for each qubit) and the $c_{i_1\dots i_n}$ are coefficients and $i_j \in \{1,2\}$. Therefore, in total we have $2^n$ number of coefficients. 
\begin{equation}
\label{eq:schmidtpsi}
\ket{\psi} =  \sum_{i}^{2^n} s_i \ket{\phi_i},
\end{equation}
where $\ket{\phi_i}$ is the $n$-qubit state whose binary index reproduces $(i_1, \dots, i_n)$ and so $\ket{\phi_i}=\ket{t_{i_1}} \otimes \dots \otimes \ket{t_{i_n}}$. 

Assuming $s_1\leq \dots \leq s_{2^n}$, similarly to $k$-rank matrix approximation, if we take only the tensor products with $k$ largest coefficients, it gives a $k$-rank tensor approximation to the vector:
\begin{equation}
\label{eq:reducedpsi}
\ket{\psi^{(k)}} = \sum_{i}^{k} s_i \ket{\phi_i}.
\end{equation}

Note that for the rest of the paper, we will assume all terms and coefficients are real in the above equations. 
\subsection{Related works}
\label{sec:relatedworks}
There are standalone quantum algorithms that implement singular value transformation \cite{gilyen2019quantum} or principal component analysis \cite{lloyd2014quantum,daskin2016obtaining}, which have been adapted to parameterized quantum circuits for face recognition problems \cite{xin2021experimental}, used for data compression \cite{yu2018quantum}, and applied to tensor‐products \cite{hastings2020classical}.

Reducing the dimension through quantum circuits are found in different machine learning models. Examples include: Ref.~\cite{andres2023efficient} where the amplitude encoding \cite{lloyd2020quantum}, linear layer preprocessing \cite{lloyd2020quantum}, and data re-uploading \cite{perez2020data} are used for dimension reduction layer for reinforcement learning. Similar techniques investigated for different datasets in Ref.~\cite{sihare2025dimensionality} and Ref.~\cite{correa2022exploring}. Ref.~\cite{liang2020variational} gives embedding strategies that preserves neighborhoods. The dimension reduction is also done by finding and using optimal projections on the dataset \cite{duan2019quantum}. Quantum circuits using random projection to lower dimension of a matrix is described in Ref.~\cite{kumaran2024random}. The coefficients in the tensor network decomposition obtained from the successive Schmidt decomposition are investigated in Ref.~\cite{daskin2023dimension} for different datasets. A quantum model described in Ref.~\cite{elliott2020extreme} for temporal data which can provide exponential compression without impeding the accuracy in the forecasting models.  A quantum resonant dimensionality reduction  algorithm that uses quantum phase estimation algorithm is proposed in Ref.~\cite{yang2025quantum}.

Quantum feature maps are shown to be universal approximators \cite{goto2021universal} and can be used successfully for supervised learning \cite{havlivcek2019supervised}. The quantum feature maps can be used to compress data \cite{dou2023efficient,matsumoto2025iterative,umeano2024ground} and can be integrated with neural networks to define hybrid graph neural networks \cite{daskin2025learnable}.

Quantum machine learning through tensor network circuits is explored in Ref.~\cite{huggins2019towards} (see the review article \cite{rieser2023tensor} for other tensor network related models). Ref.~\cite{kardashin2021quantum} describes a quantum algorithm for returning a classical description of a tensor network state and approximating an eigenvector. The tree structure of the tensor network of the datavector is used in Ref.~\cite{wall2021tree} for assisting the classifier. The tensor networks are also used to define interpretable quantum machine learning models \cite{ran2023tensor}.

\section{Methods}
\label{sec:methods}
In the linear combination of tensor products given in Eq.~\eqref{eq:schmidtpsi}, each term $\ket{\phi_i}=\ket{t_{i_1}} \otimes \dots \otimes \ket{t_{i_n}}$ can be implemented on quantum circuits by using single-qubit $R_y$ rotation gates on each qubit.  The linear combination  of these circuits  can be done by using block encoding \cite{childs2012hamiltonian,daskin2012universal} with the utilization of an ancilla register. In this circuit, a register with $\lceil \log_2 k \rceil$ ancilla qubits is used to control $k$ distinct terms. And the coefficients are implemented applying a layer of rotation gates ($R_y$) and controlled $R_y$ gates to the ancilla register before the controlled operations of the terms.  

For a dataset $\{\vec{x}_1, \dots, \vec{x}_m\}$ where each $\vec{x}_i \in \mathbb{R}^{2^n}$, our objective is to construct a quantum circuit $U$ such that for any input state $\ket{\vec{x}_i}$, the circuit generates its reduced form $\ket{\vec{x}_i^{(k)}}$. 
To achieve this, we will design a circuit $U(\vec{\theta})$ with learnable parameters $\vec{\theta}$ that implements the $k$ term linear combination of the tensor products specified in Eq.\eqref{eq:reducedpsi}. 

The main algorithmic steps to determine $k$ and implement circuit $U$ are as follows:
\begin{enumerate}
    \item  We first compute the mean data vector from a given training sample: 
\begin{equation}
    \vec{\bar{x}} = \frac{\sum_i^m \vec{x}_i}{{\|\sum_i^m \vec{x}_i}\|_2}.
\end{equation}
\item Then, compute its tensor decomposition by using successive Schmidt decompositions \cite{daskin2023dimension}:
\begin{equation}
    \vec{\bar{x}} = \ket{\psi} = \sum_{i}^{2^n} s_i \ket{\phi_i}.
\end{equation}
\item Then, apply coefficient thresholding ($s_i \geq \gamma$) to obtain a $k$-term approximation :
\begin{equation}
    \vec{\bar{x}^{(k)}} = \ket{\psi^{(k)}}=\sum_{i}^{k} s_i \ket{\phi_i} \quad \text{ with } s_i \geq \gamma.
\end{equation}
Here, note that since the threshold is used as the parameter to determine the value of $k$ when it is too small the value for $k$ maybe exponentially large. Therefore, a maximum value for a $k$ can be used to prevent such cases.
\item For real-valued vectors and coefficients, implement each $i$-th term using $R_y$ rotations:
\begin{equation}
R_y{(\theta_{i1})} \otimes R_y{(\theta_{i2})} \otimes \dots \otimes R_y{(\theta_{in})}      
\end{equation}
For complex vectors a generic unitary gate can be formed by using Euler decomposition \cite{nielsen2010quantum} with $R_z$ and $R_y$ gates.
\item Implement the linear combination via block encoding: 
    \begin{itemize}
        \item Allocate $\lceil \log_2 k \rceil$ ancilla qubits
        \item Apply rotation and entangling gates ($R_y$ and controlled-$R_y$) to ancilla
        \item Control each tensor product term using unique ancilla basis states
    \end{itemize}
The resulting circuit is depicted in Fig.\ref{fig:circuit} for $k=3$ where we have 2 ancilla qubits, 4 data qubits, 2 layer of rotation and entangling gates on the ancilla and 3 terms of product implementation.

\item \textbf{Measurement note:} The final measurement combines controlled unitaries without additional ancilla rotations. Therefore, we did not use another layer of rotation gates on the ancilla register after the controlled operation. If the circuit is used as a preprocessing gadget for another quantum circuit, another layer of rotation gates is necessary on the ancilla to have the linear combination of the products.

\begin{figure}[ht]
    \centering
    \includegraphics[width=1\linewidth]{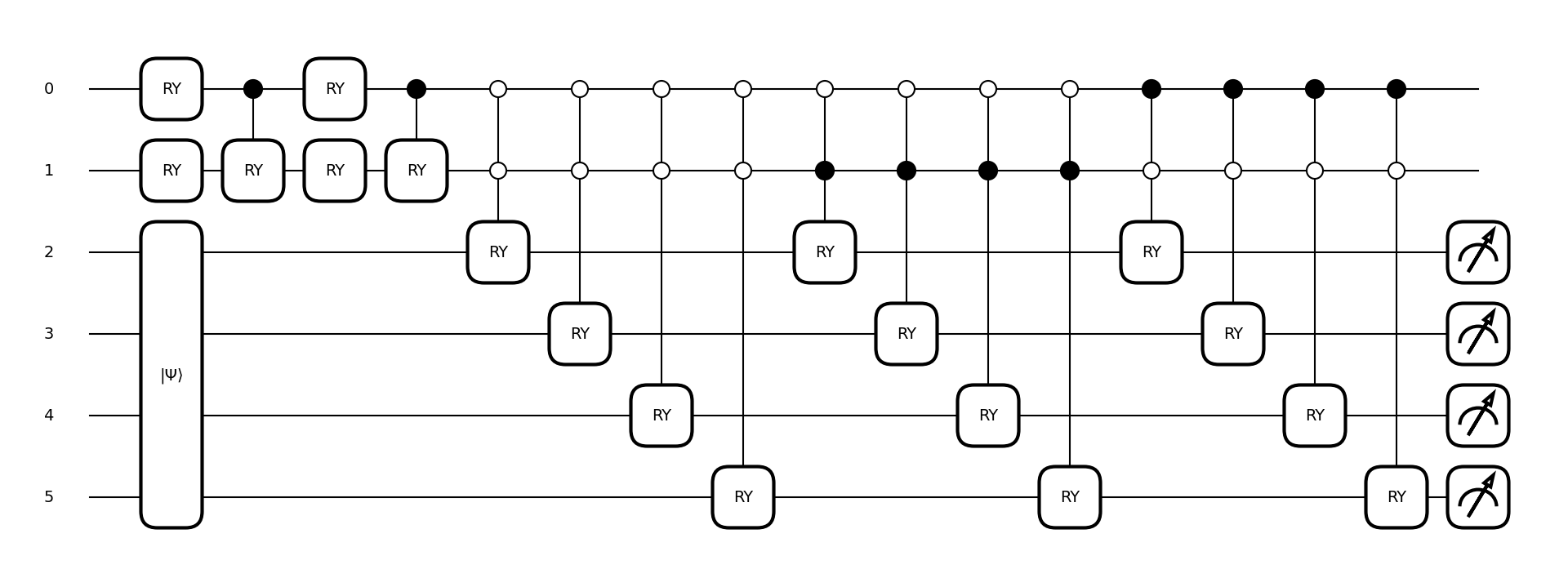}
    \caption{Example circuit for the linear combination of three tensor‐product terms: the upper two control qubits form an ancilla register and have two layers of circuit ansatz. The measurement is performed only on the data register.}
    \label{fig:circuit}
\end{figure}

\end{enumerate}
\subsection{Optimizing circuit parameters} 
We optimize the learnable parameters of the circuit for a given data set by executing circuit for each data vector $\vec{x}_i$, measuring the output probabilities, and adjusting its parameters through a classical optimization routine. 

\subsection{Use in machine learning} 
The probabilities obtained at the end of the circuit is of dimension $n$. Therefore, it provides an exponential compression of the data vectors. These probability vectors can serve as inputs to classical processing components, which may include: neural network heads, traditional other machine learning models or optimization frameworks.
Note that parameter optimization can be performed either as preprocessing or jointly in combination of the machine learning framework to construct an end-to-end machine learning pipeline. The latter is used as a hybrid machine learning model in the experiments section: Section~\ref{sec-hybrid}.

\section{Experiments and numerical analyses}
\label{sec:experiments}
\subsection{How well real world datasets can be approximated}
\label{sec:how-well}
\begin{figure*}
    \centering
    \includegraphics[width=1\linewidth]{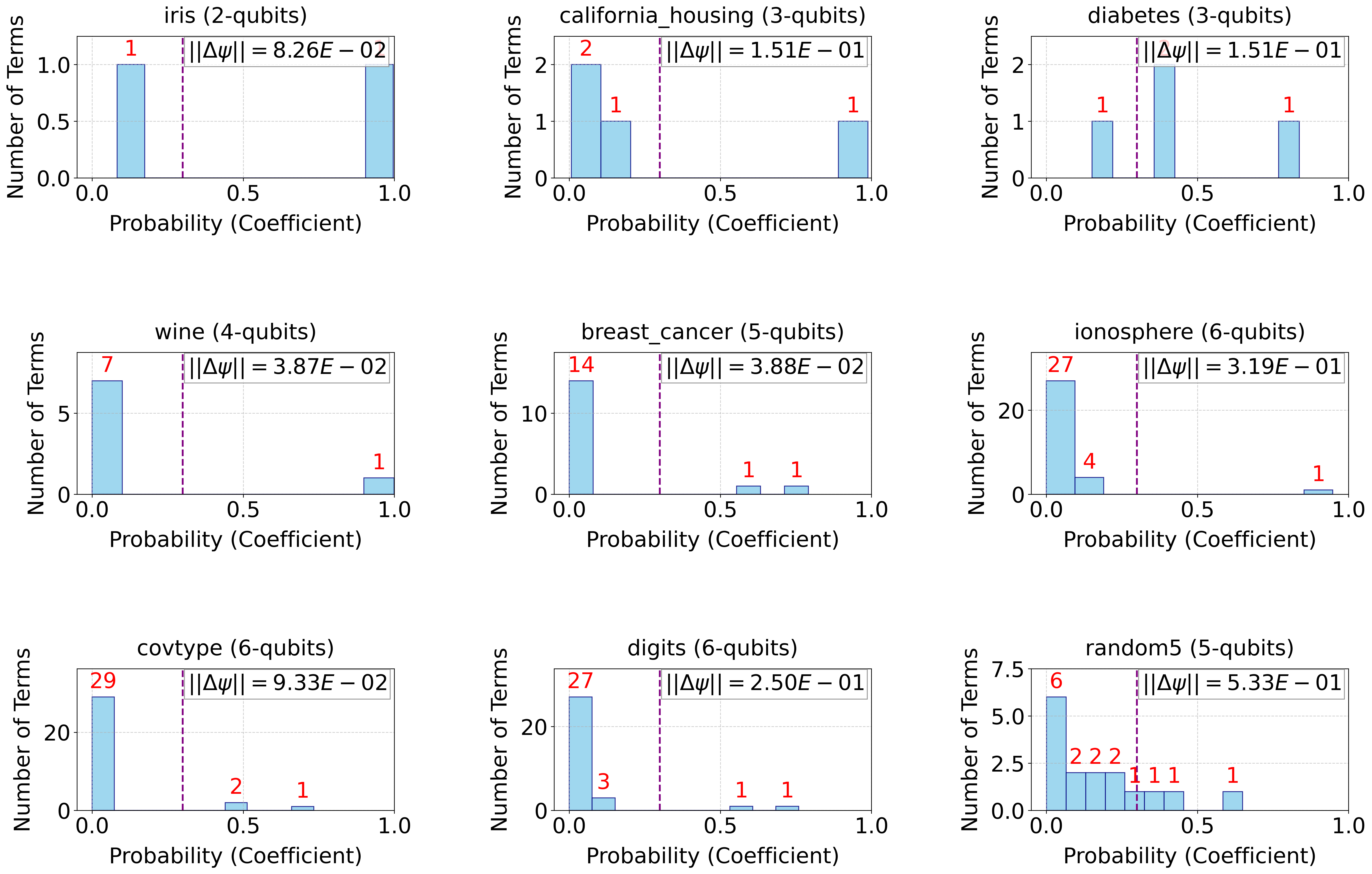}
    \caption{The coefficient distributions for datasets requiring $\leq$6 qubits. Vertical dashed line indicates threshold $\gamma=0.3$; $\|\Delta\psi\|$ denotes approximation error. Decompositions computed for mean vectors of 100-sample subsets. ``random5'': synthetic dataset generated via \texttt{make\_classification} for 5-qubit classification.}
    \label{fig:sd-small}
\end{figure*}

\begin{figure*}
    \centering
    \includegraphics[width=1\linewidth]{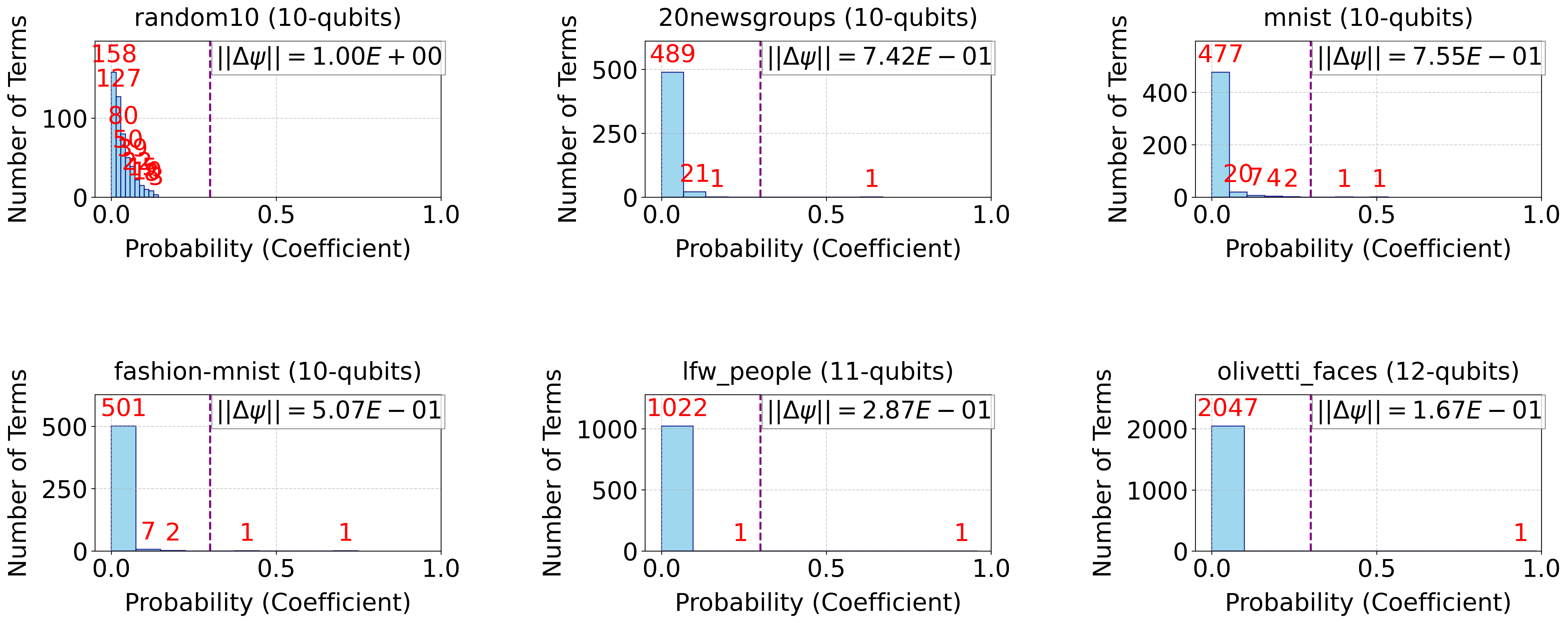}
    \caption{The coefficient distributions for datasets requiring $>$6 qubits. Vertical dashed line indicates threshold $\gamma=0.3$; $\|\Delta\psi\|$ denotes approximation error. Decompositions computed for mean vectors of 100-sample subsets. ``random10'': synthetic dataset generated via \texttt{make\_classification} for 10-qubit classification.}
    \label{fig:sd-large}
\end{figure*}

The coefficients in the tensor network decomposition have been analyzed for many datasets in Ref.\cite{daskin2023dimension} where they primarily consider $\text{vec}(XX^T)$ for the data matrix $X$. 
For completeness, here we present similar coefficient analyses for datasets accessible via Python's scikit-learn module \cite{scikit-learn}. 
Figures~\ref{fig:sd-small} and~\ref{fig:sd-large} display histograms of the coefficient  values (stated as probability since found through Schmidt decompositions) for mean vectors across different dataset samples.  From these figures we can see that many data sets can be well-approximated by using a cut-off value for the threshold. In the figures, the vertical dashed lines indicate our chosen threshold $\gamma = 0.3$, with corresponding approximation errors annotated by $\|\Delta\psi\|$.
We note that the observed gaps in the probabilities are not specific to these datasets. Many datasets show similar properties. 

\subsection{$k$-rank approximation via quantum circuit}
\label{sec:approximation-through-qc}
We employ the quantum circuit shown in Fig.~\ref{fig:circuit} with $k=3$ for relatively small datasets, using the same threshold value $\gamma=0.3$ as in Fig.~\ref{fig:sd-small}.

During optimization, we compare the $2^n$-dimensional output probability vector from the data register with target values obtained via successive Schmidt decomposition. Note that when we use this circuit in the hybrid machine learning model, this circuit provides $n$-dimensional probability outputs instead of $2^n$-dimensional state vectors, enabling exponential data dimensionality reduction.

\begin{figure}[ht]
    \centering
    \includegraphics[width=1\linewidth]{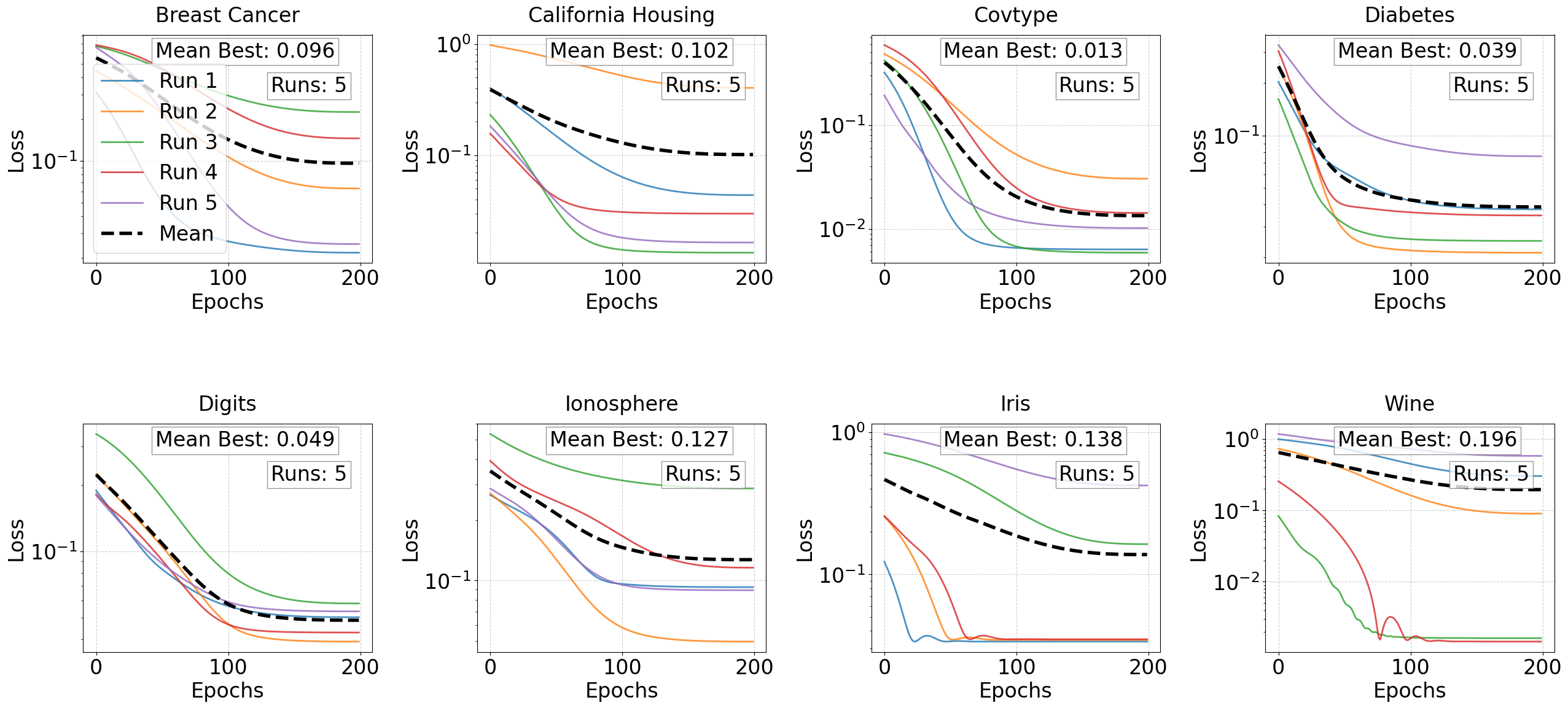}
    \caption{Optimization results for individual vectors: Loss represents $\|P_{\text{obtained}} - P_{\text{target}}\|_F$ between circuit outputs and classically precomputed targets. Results shown for 5 optimization runs on single vectors randomly selected from 20-sample subsets. ``Mean Best'' is the average minimum loss across runs.}
    \label{fig:sdcircuit-meanvector}
\end{figure}

\begin{figure}[ht]
    \centering
    \includegraphics[width=1\linewidth]{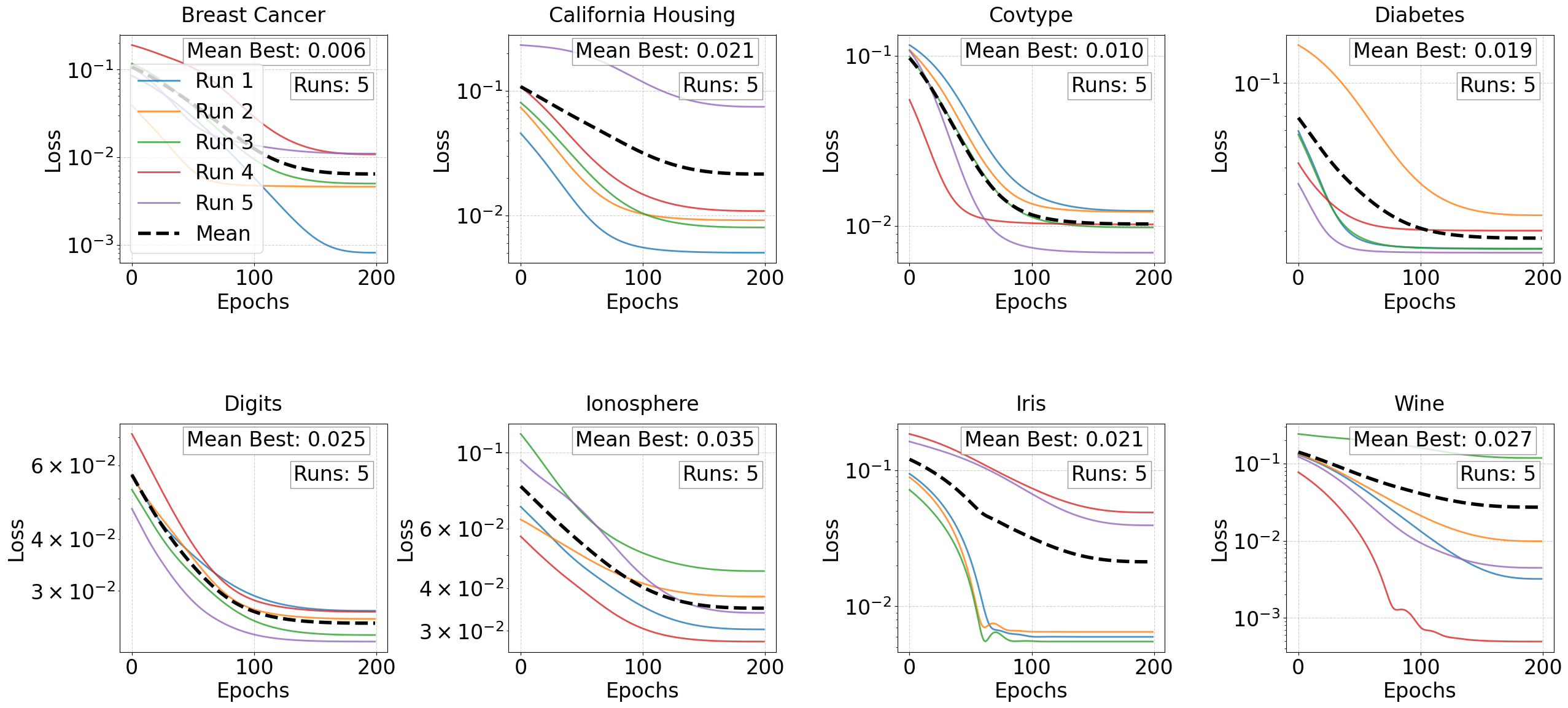}
    \caption{Optimization results for full samples: Loss represents $\frac{1}{20}\sum_i^{20}\|P_{\text{obtained}}^{(i)} - P_{\text{target}_i}\|_F$ averaged across 20-vector samples. Results shown for 5 optimization runs per dataset. ``Mean Best'' is the average minimum loss across runs.}
    \label{fig:sdcircuit-sample}
\end{figure}

The optimization results are presented in Fig.~\ref{fig:sdcircuit-meanvector} for individual vectors and Fig.~\ref{fig:sdcircuit-sample} for full samples.  The loss function is defined as the norm of the difference between the output probabilities and the target probabilities (for a single vector) or the average of such norms (for the whole sample):
\begin{equation}
    Loss = \frac{\sum_i^{m_s} \|P_{obtained} - P_{target_i}\|_F}{m_s},
\end{equation}
where $\|.\|_F$ represents the Frobenius norm of vectors, $m_s$ is the sample size, $P_{obtained}$ and $P_{target_i}$ are  $2^n$ dimensional probability vectors obtained from the quantum circuit and the computed classically in pre-processing for each vector $\vec{x}_i$ int the sample.
The value of $k$ and number of qubits used in the approximation is given in Table \ref{tab:dataset-parameters}.
\begin{table}[h]
\centering
\caption{The number fo parameters used in the circuit optimization for the dataset samples.}
\label{tab:dataset-parameters}
\resizebox{0.9\textwidth}{!}
{
\begin{tabular}{@{}l|c|c|c@{}}
\toprule
\textbf{Dataset}    & \textbf{Number of Qubits} & \textbf{Number of Terms (k)} & \textbf{Number of Ancilla Qubits} \\
\hline
iris                & 2                         & 1                            & 0                                 \\
california\_housing & 3                         & 1                            & 0                                 \\
diabetes            & 3                         & 3 or 2                       & 2 or 1                            \\
wine                & 4                        & 1                            &  0                                 \\
breast\_cancer      & 5                         & 2                            & 1                                 \\
ionosphere          & 6                         & 1 or 2                       & 0 or 1                            \\
covtype             & 6                         & 3                            & 2                                 \\
digits              & 6                         & 2                            & 1     \\
\bottomrule
\end{tabular}
}
\end{table}
We employ classical AdamW optimizer from Python's PyTorch module with a decaying learning rate of 0.05. The optimization results are presented in Fig.~\ref{fig:sdcircuit-meanvector} for single vectors and in Fig.~\ref{fig:sdcircuit-sample} for samples of  20 data vectors. The optimization is run 5 times for each dataset with a randomly chosen sample. Mean best loss values are also noted in the figures.

Comparing the results in Fig.~\ref{fig:sdcircuit-meanvector} and  Fig.~\ref{fig:sdcircuit-sample} with the classically obtained approximation errors noted in Fig.~\ref{fig:sd-small}, it can be seen that the results obtained from quantum circuits are comparable and in some cases better. This indicates that the circuit can be used to obtain a $k$-rank approximation of a dataset.

\subsection{Comparison of reduced vectors to original data in classical machine learning}
\label{sec:comparison-in-classical}
\begin{figure}[ht]
    \centering
    \includegraphics[width=1\linewidth]{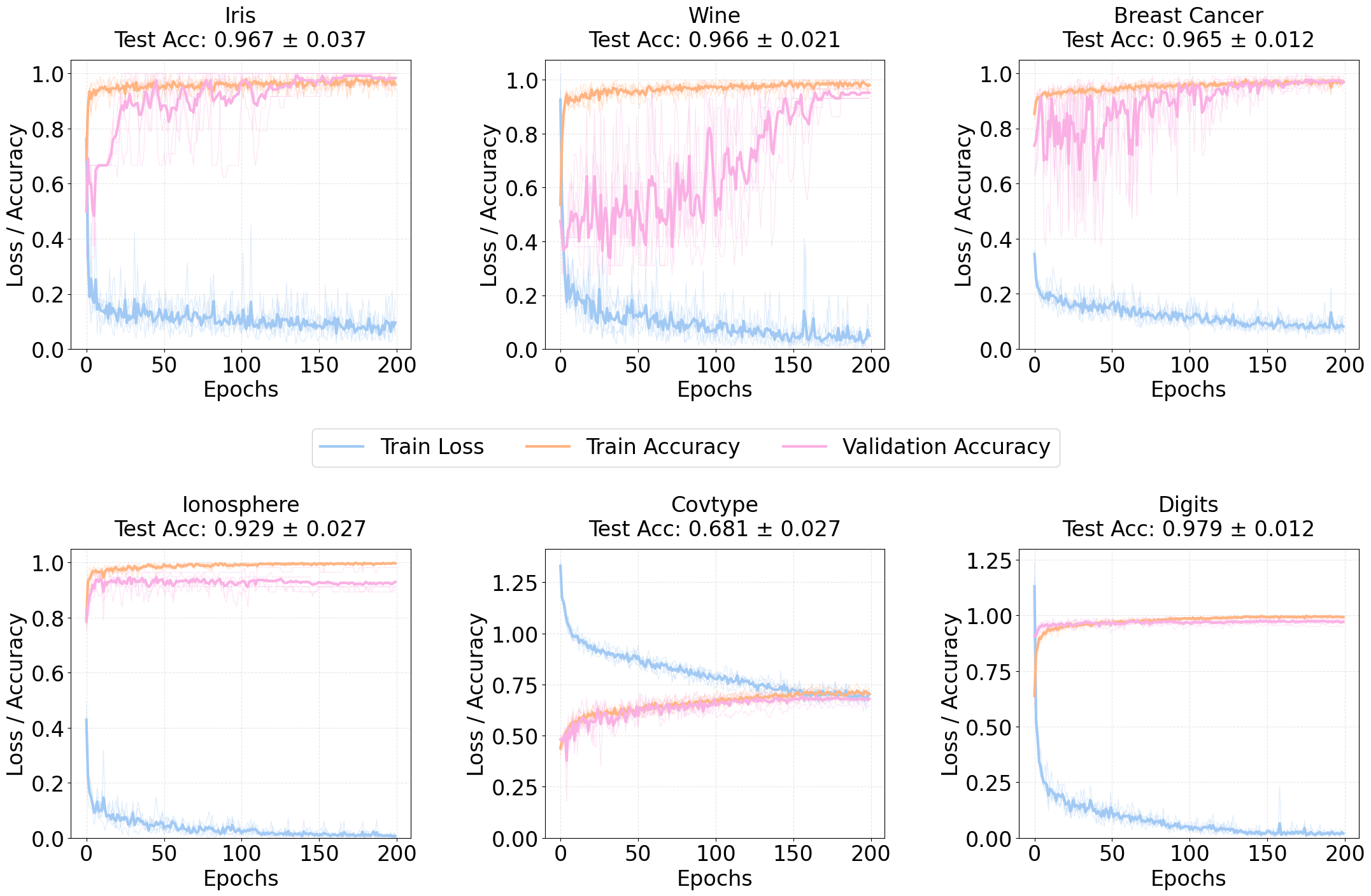}
    \caption{5-fold cross validation: Classical neural network performance on original (unreduced) datasets. Maximum sample size is 1000.}
    \label{fig:classical-full}
\end{figure}

\begin{figure}[ht]
    \centering
    \includegraphics[width=1\linewidth]{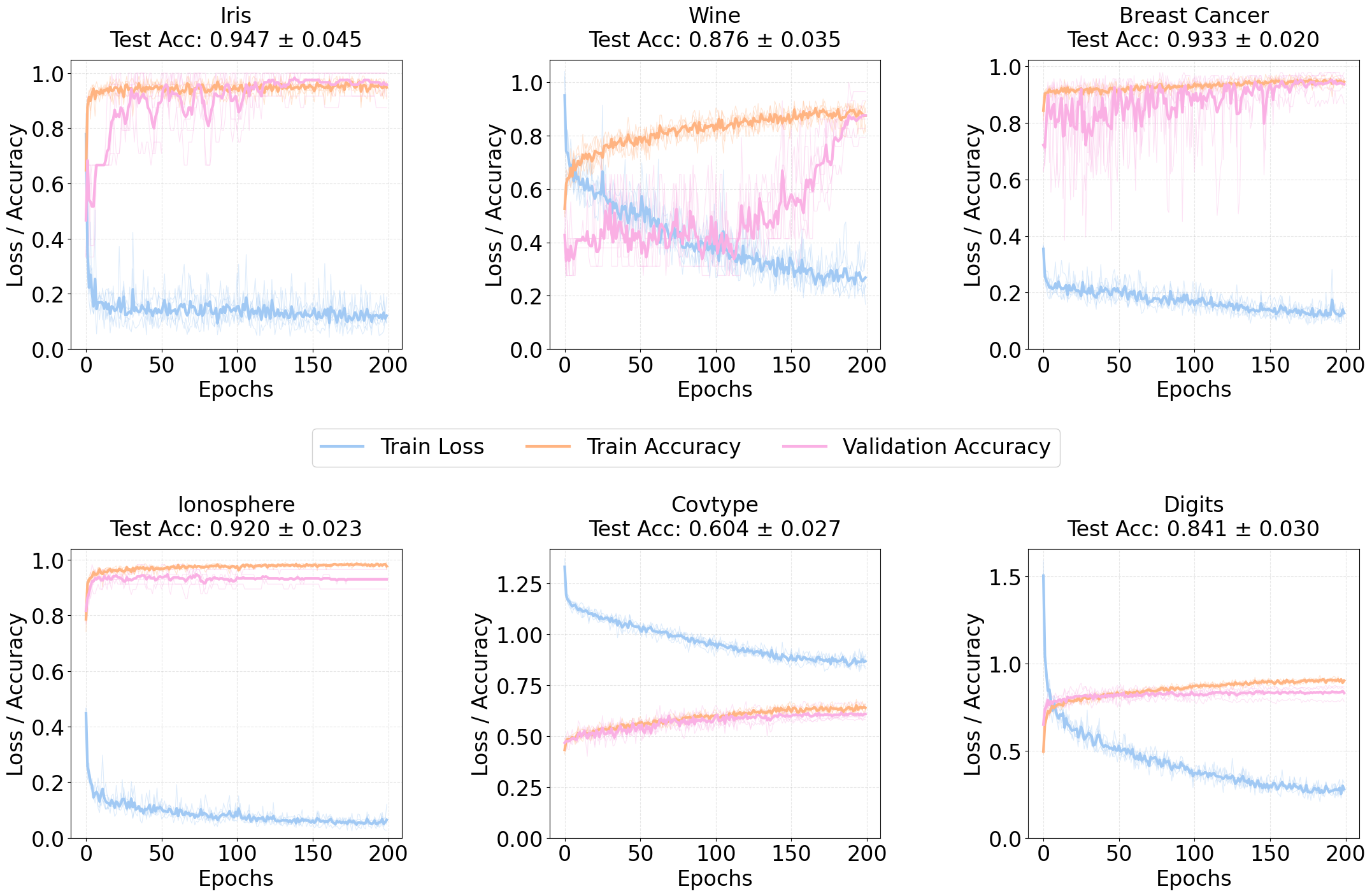}
    \caption{5-fold cross validation: Classical neural network performance on reduced ($k$-rank) datasets. Maximum sample size is 1000.}
    \label{fig:classical-reduced}
\end{figure}

While $\|\Delta\psi\| = \|\ket{\psi}-\ket{\psi}^{(k)}\|$ quantifies the approximation error for individual vectors, from this value alone, it is difficult to directly predict machine learning performance when substituting original vectors $\{\vec{x}_1,\dots,\vec{x}_m\}$ with reduced versions $\{\vec{x}_1^{(k)},\dots,\vec{x}_m^{(k)}\}$. To evaluate this performance impact, we implement a classical neural network head in PyTorch (without quantum components) for both data representations. The architecture is specified as:

\begin{center}
\begin{minipage}{0.8\textwidth}
\begin{verbatim}
(classical_net):
(0): Linear(in_features=2^n, out_features=64, bias=True)
(1): BatchNorm1d(64, eps=1e-05, momentum=0.1, affine=True, track_running_stats=True)
(2): ReLU()
(3): Dropout(p=0.25, inplace=False)
(4): Linear(in_features=64, out_features=32, bias=True)
(5): BatchNorm1d(32, eps=1e-05, momentum=0.1, affine=True, track_running_stats=True)
(6): ReLU()
(7): Dropout(p=0.25, inplace=False)
(8): Linear(in_features=32, out_features=n_classes, bias=True)
\end{verbatim}
\end{minipage}
\end{center}
Figures~\ref{fig:classical-full} and~\ref{fig:classical-reduced} present learning curves and final accuracies for both original and reduced datasets using 5-fold cross validation, with maximum sample sizes capped at 1000 per dataset.
We can see from the figures that accuracy degradation due to dimensionality reduction is significant for ``Covtype", ``Digits" and ``Wine" datasets while the others are less than $\approx 5\%$. 

\subsection{Hybrid machine learning model}
\label{sec-hybrid}
The quantum circuit described in Section~\ref{sec:methods} can function either as a standalone preprocessing module or as an integrated component with a classical neural network head. In this work, we integrate it with the classical architecture specified in Section~\ref{sec:comparison-in-classical}, with one critical modification: the input dimension to the classical head reduces from $2^n$ to $n$.

This means quantum circuit not only reduces dimension but also exponentially compresses it which dramatically decreases learnable parameters (see Table~\ref{tab:parameters}). This combined effect is particularly significant for larger datasets, where classical layer sizes would otherwise scale exponentially. 
\begin{table}[h]
\centering
\caption{Numbers of learnable parameters in the hybrid model with standalone classical head applied to data vectors with and without dimension reduction.}
\label{tab:parameters}
\resizebox{0.9\textwidth}{!}
{
\begin{tabular}{@{}l|c|c|c@{}}
\toprule\\ 
\textbf{Dataset} & \textbf{Hybrid Model} & \textbf{Classical Head with Reduced Vectors} & \textbf{Classical Head with Original Vectors} \\
 \hline 
 iris             & \textbf{2565}                             & 2693                                                             & 2693                                          \\
wine             & \textbf{2695}                             & 3463                                                             & 3463                                          \\
breast\_cancer   & \textbf{2734}                             & 4462                                                             & 4462                                          \\
ionosphere       & \textbf{2792}                             & 6504                                                             & 6504                                          \\
covtype          & \textbf{2975}                             & 6687                                                             & 6687                                          \\
digits           & \textbf{3064}                             & 6776                                                             & 6776                                          \\ \bottomrule
\end{tabular}
}
\end{table}

\begin{figure}[ht]
    \centering
    \includegraphics[width=1\linewidth]{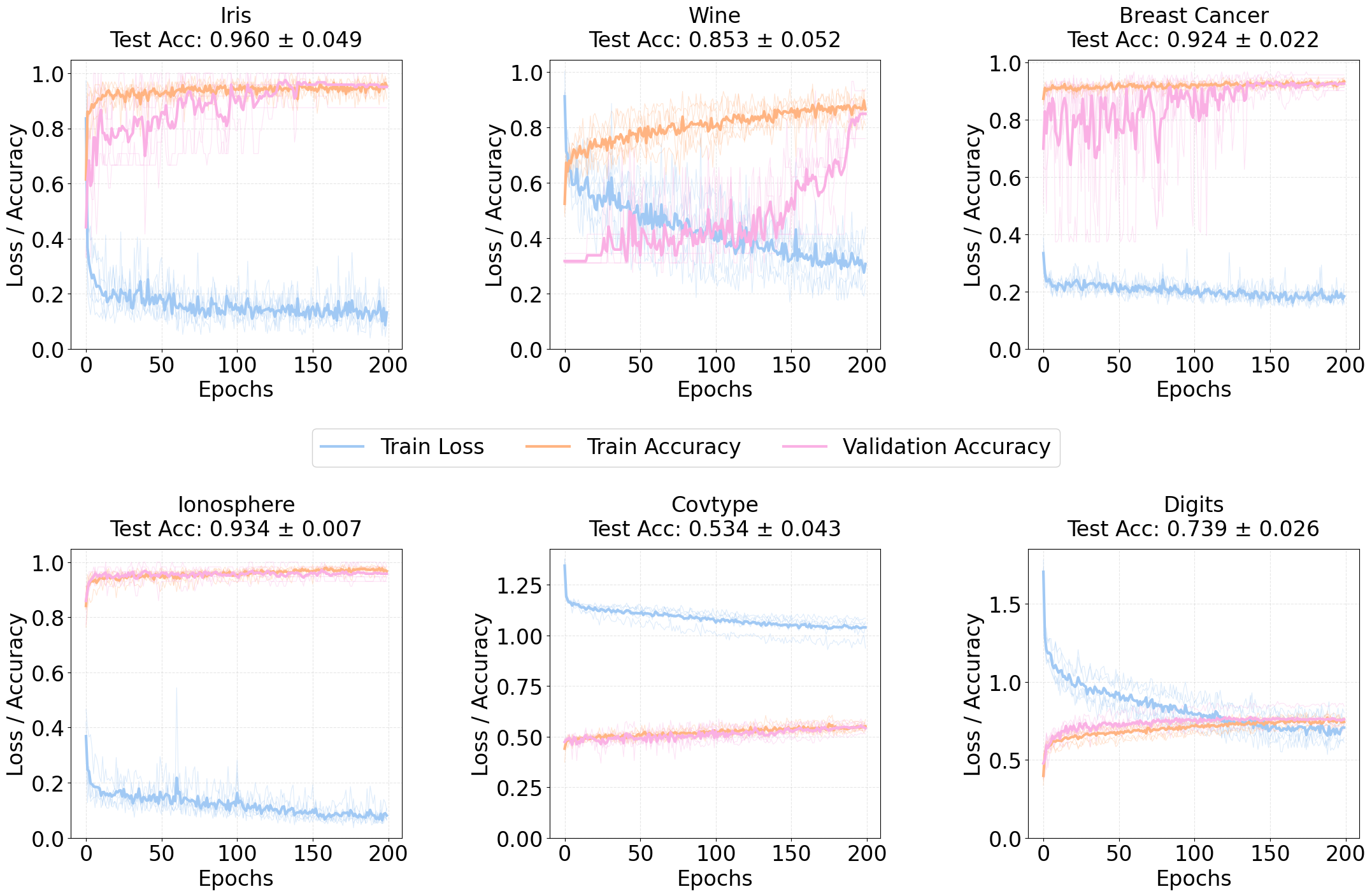}
    \caption{5-fold cross validation results for hybrid quantum-classical model. The quantum circuit component provides exponential dimensionality reduction of input features (from $2^n$ to $n$ dimensions) for the classical neural network head. Maximum sample size is 1000.}
    \label{fig:cv-hybridmodel}
\end{figure}

The success of the proposed model is evaluated by comparing its performance against the classical head applied to reduced datasets (Fig.~\ref{fig:classical-reduced}). Comparable results would confirm that the quantum circuit achieves exponential dimensionality reduction while preserving essential information.
Using identical optimization settings, we report hybrid model results in Fig.~\ref{fig:cv-hybridmodel}. For direct comparison, Table~\ref{tab:results} summarizes key findings: despite exponential compression, the hybrid model achieves performance comparable to—and in some cases superior to—the classical approach operating solely on reduced datasets.

\begin{table}[]
\centering
\caption{Comparison of the hybrid model with standalone classical head applied to data vectors with and without dimension reduction through successive Schmidt decompositions.}
\label{tab:results}
\resizebox{0.9\textwidth}{!}
{
\begin{tabular}{@{}l|c|c|c@{}}
\toprule\\
\textbf{Dataset} & \textbf{Hybrid Model} & \textbf{Classical Head with Reduced Vectors} & \textbf{Classical Head with Original Vectors} \\ \hline 
iris             & 0.960 ± 0.049                             & 0.947 ± 0.045                                                    & \textbf{0.967 ± 0.037}                                            \\
wine             & 0.853 ± 0.052                             & 0.876 ± 0.035                                                    & \textbf{0.966 ± 0.021}                                            \\
breast\_cancer   & 0.924 ± 0.022                             & 0.933 ± 0.020                                                    & \textbf{0.965 ± 0.012}                                            \\
ionosphere       & \textbf{0.934 ± 0.007}                    & 0.920 ± 0.023                                                    & 0.929 ± 0.027                                                     \\
covtype          & 0.534 ± 0.043                             & 0.604 ± 0.027                                                    & 0.681 ± 0.027                                                     \\
digits           & 0.739 ± 0.026                             & 0.841 ± 0.030                                                    & 0.979 ± 0.012                                                     \\ \bottomrule
\end{tabular}
}
\end{table}

\section{Discussion and conclusion}
\label{sec:discussion}
This paper introduces tensor network structure-aware quantum circuits for hybrid machine learning frameworks. By leveraging dataset-specific tensor network decompositions, our quantum circuits dramatically reduce learnable parameters while maintaining competitive accuracy when combined with classical neural network heads.

The circuit model employs $k$ terms determined from the mean vector of training sample. While this approach risks overfitting to training data characteristics, it can provide critical advantages such as preventing quantum circuit over-parameterization, mitigating barren plateau problems, and enabling interpretability through measurable approximation error ($\|\Delta\psi\|$).
A possible future work should explore alternative $k$-selection heuristics beyond sample mean.

As final note, we should note that in our hybrid model we train the quantum part and the classical head together. This can lead complications for the deployment of trained models since it may require connection to quantum APIs. 
\section{Data availability}
The simulation code and results used for this paper are publicly available
at: \url{https://github.com/adaskin/structure-aware-circuits}
\section{Funding}
This project is not funded by any funding agency.

\section{Conflict of interest}
The authors declare no conflict of interest.

\bibliographystyle{unsrt}
\bibliography{main}
\end{document}